 \documentclass[12pt,preprint]{aastex}

\newcommand{\vm}{$V_{max}$}
\newcommand{\ival}{$I_{val}$}

\slugcomment{Submitted to the Astronomical Journal}

\shorttitle{A Catalog of Contact Binary Stars}
\shortauthors{Gettel et al.}

\begin{document}


\title{A Catalog of 1022 Bright Contact Binary Stars}


\author{S. J. Gettel\altaffilmark{1,2}, M. T. Geske\altaffilmark{1} and T. A. McKay\altaffilmark{1}}


\altaffiltext{1}{Department of Physics, University of Michigan,
    Ann Arbor, MI 48109}
\altaffiltext{2}{sjgttl@umich.edu}


\begin{abstract}
In this work we describe a large new sample of contact binary stars extracted
in a uniform manner from sky patrol data taken by the ROTSE-I telescope. Extensive ROTSE-I
light curve data is combined with J, H, and K band near-infrared data taken from the Two
Micron All-Sky Survey (2MASS) to add color information. Contact binaries candidates are
selected using the observed period-color relation. Candidates are confirmed by visual 
examination of the light curves. To enhance the utility of this catalog, we derive a 
new {\sl J--H} period-color-luminosity relation and use this to estimate distances for
the entire catalog. From these distance estimates we derive an estimated contact binary 
space density of $1.7 \pm 0.6 \times 10^{-5} pcs^{-3}$.
\end{abstract}




\section{INTRODUCTION}

W UMa-type contact binaries are eclipsing systems in which both stars overflow their Roche lobes, 
forming a common envelope of material \citep{luc68}. They are formed from nearly normal main-sequence 
stars with spectral types usually between F and K. Typical mass ratios are between $q 
\approx 0.2 - 0.5$, but reported values are almost as high as unity, and as low as 0.066 \citep{ruc01}. 
Their periods range from 0.22 days to 1.5 days, with most systems having periods between $0.25 - 0.5$ days. 
Because of their close proximity, they produce continuously varying light curves, making them detectable 
at a large range of inclinations. 

Due to the common atmosphere, both stars have essentially the same surface temperature. Through a 
combination of Kepler's third law and the radius-color relationship for main-sequence stars, this 
common temperature leads to a period-color-luminosity relation (\citet{ruc94}, \citet{ruc97}). This 
relation, along with their ease of detection, make contact binaries useful tracers of distance and 
galactic structure, especially on small scales \citep{ruc04}. 

Contact binaries are known to be quite common among variable stars, though their space density is 
still debated. Early estimates range from 10$^{-6}$ pc$^{-3}$ \citep{kra67} to 10$^{-4}$ pc$^{-3}$ 
\citep{van75}. More recent estimates include those of \citet{ruc02}, who found a density of 
$1.0 \times 10^{-5} pc^{-3}$, or $1/500$ main-sequence stars, in the solar neighborhood. 
This conflicts with a previous estimate of $1/130$ main-sequence stars made using OGLE-I 
data in the galactic disk \citep{ruc98b}, and may be an indication of significant variability in 
the contact binary fraction through the galaxy.

Currently, the GCVS \citep{sam04} labels 845 stars as EW type variables; W UMa variables with 
periods less than 1 day and nearly equal minima. However, GCVS classifications 
are not entirely secure. Other contact binary catalogs include that of \citet{pri03}, which 
contains 361 galactic EW and EB type variables, each previously 
identified by another catalog such as the GCVS. Contact binaries included had 
either light curve or good spectroscopic data available. Also, analysis of the ROTSE-I 
variability test fields \citep{ake00a} found 382 contact binary candidates within about 2000 deg$^{2}$, 
identified by their light curve shape and period. 
Periods and light curve data were provided for all of these ROTSE-I objects.

We present here a catalog of 1022 contact binary stars, 836 of which are not found in the GCVS, SIMBAD 
database, Pribulla catalog or ROTSE-I test fields. Light curve data, periods, and distance estimates 
are presented for each object. These objects passed a rather stringent selection process and have high quality 
light curves, so their classification is relatively secure. The completeness of this catalog for the 
regions surveyed is estimated to be about 34\%, with the remaining objects lost to data quality cuts. 
Near infrared observations of these objects drawn from the Two-Micron All-Sky Survey (2MASS) allow us 
to create a NIR period-color-luminosity relation and to estimate distances to each object. Finally, an 
estimate of the space density of these objects is made. 
  
Section \ref{sec:catalog} contains an overview of the data and selection methods, 
which is followed by a description of the method used to obtain distance estimates in Section 
\ref{sec:distances}. The space density derived from this catalog 
is discussed in Section \ref{sec:space_density}. Section \ref{sec:conclusions} contains a summary of the results. 

\section{ASSEMBLING THE CONTACT BINARY CATALOG \label{sec:catalog}}

In this section we describe how we assemble a catalog of bright contact binary
stars for study. We begin with a short description of the ROTSE-I instrument,
which obtained the data, and the NSVS database on which this project is 
based. This is followed by a description of variable selection, identification
of short period variables, phasing of these variables, combination of
ROTSE-I optical data with 2MASS near-infrared data, and finally the selection
of contact binaries from this catalog.

\subsection{NSVS Data}

Optical variability data were obtained by the ROTSE-I robotic 
telescope\footnote{For more information about ROTSE see http://www.rotse.net}.
ROTSE-I was a four-fold array of Canon 200 mm f/1.8 lenses, each equipped
with an unfiltered 2048x2048 pixel Thompson TH7899M CCD. At this f-number, the 14 $\mu$m 
pixels of the CCD subtend 14.4\arcsec\ on the sky. The combined array imaged
a continuous 16$^\circ$x16$^\circ$ field of view. ROTSE-I, designed to pursue 
real-time observations of gamma-ray bursts, spent most of the time from March 
1998 until December 2001 patrolling the sky. The very large ROTSE-I field of
view allowed it to image the entire available sky twice each night, taking 
a pair of 80 s images during each visit. Typical limiting magnitudes range
from $m_{v}=14.5-15.5$, depending on sky conditions. During its operating 
life, ROTSE-I amassed a 7 terabyte time domain survey of the night sky. Since 
being disassembled in 2002, the ROTSE-I lens and camera assemblies have been 
reborn as elements of the Hungarian Automated Telescope Network (HAT-net; 
\citet{bak04}).

Initial studies of ROTSE-I sky patrols were reported in \citet{ake00a}. 
Though this work examined only three months of observations covering just 5\% of 
the sky patrol area it revealed nearly 1800 bright variable objects, most of which 
were previously unknown. More recently, light curve data from a full year of all 
ROTSE-I sky patrols has been released publicly by \citet{woz04} as the Northern 
Sky Variability Survey (NSVS), which is available through 
the SKYDOT website at Los Alamos National Lab (skydot.lanl.gov). This work 
includes the entire region north of -30$^\circ$ declination, though 
coverage is neither perfectly uniform nor absolutely complete. Many more light 
curve points are available for sources at high declination than at low. Completeness is 
reduced in regions of very high stellar density.

All optical light curves used in this paper are drawn from the NSVS. Details of
the SExtractor \citep{ber96} based reductions and relative photometry corrections 
of ROTSE-I data for inclusion in the NSVS catalog, along with maps of source
density and number of good light curve data points, are presented in 
\citet{woz04}.

\subsection{Selection and Phasing of Variable Objects with Short Periods}

Selection of variable objects from the NSVS light curve database follows the
methods outlined in \citet{ake00a}. For each object, available data are 
examined for all good coincident pairs of observations. For this purpose,
`good' points were defined at two levels, one more tolerant than the other.
Cuts are made by examining the measurement flags described in \citet{woz04}.
Our tight cuts require no processing flags except the SExtractor `blended' 
flag, indicating that an object is the result of a deblending procedure. In 
addition to the `blended' flag, our second set of loose cuts allow the 
inclusion of points that have `nocorr' and `patch' flags, indicating that the 
relative photometry correction of an object could not be estimated, or that the map of corrections was patched to obtain the value for that object. 
Both sets of cuts are quite restrictive and leave us with a set of very
well measured light curves. 

\begin{figure}
\epsscale{1}
\plotone{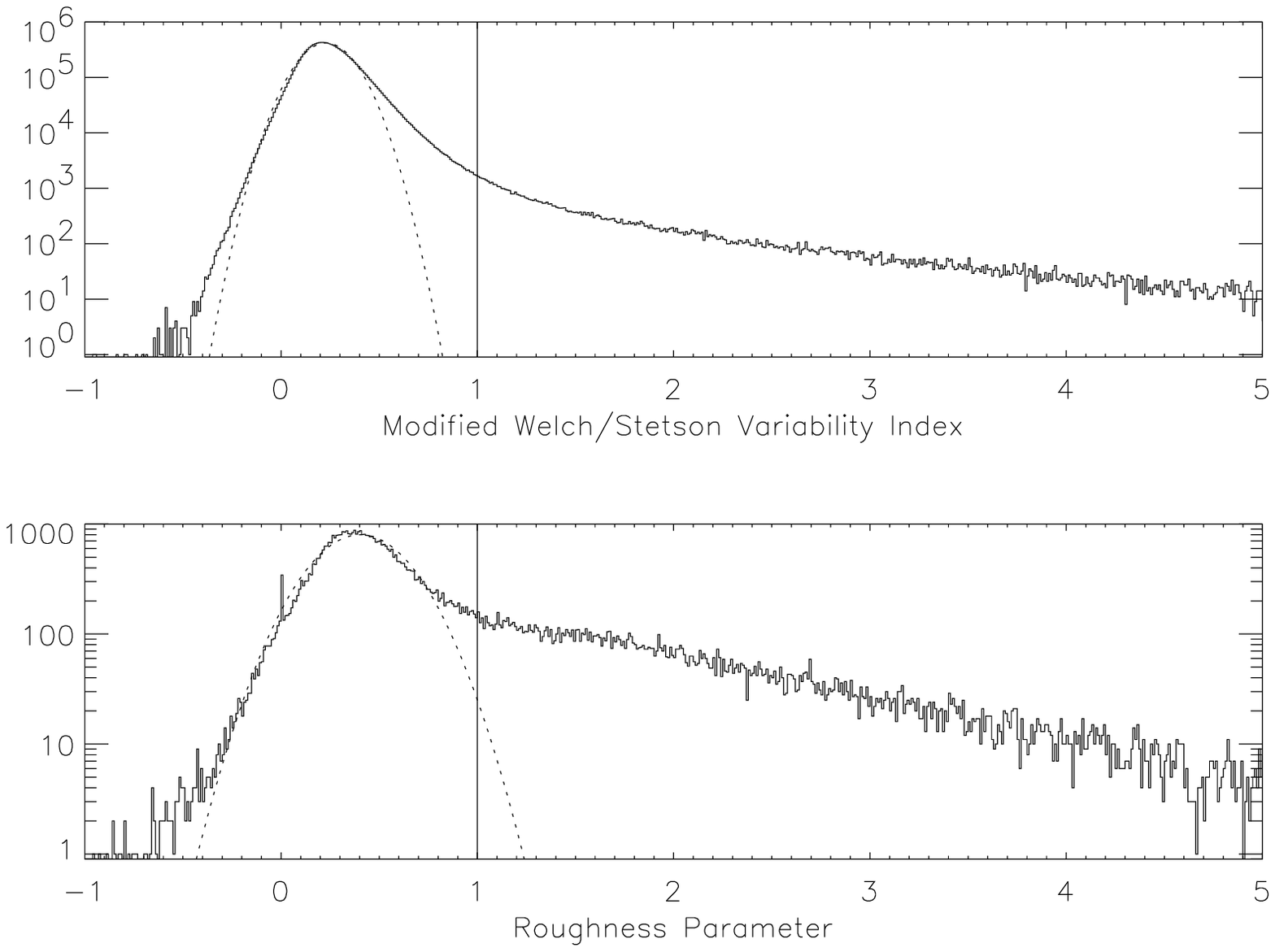}
\figcaption[figure1.eps]{The figure on the top shows 
the distribution 
of modified Welch/Stetson variability indices for a representative subset of
NSVS data. The dotted line shows a Gaussian fit to this distribution, and the
solid line shows the variability cut (about seven $\sigma$) applied in forming
this catalog. The bottom figure shows the distribution of
the roughness parameter R described in the text for the 63665 variables 
extracted from the NSVS light curve database. The dotted line shows a 
Gaussian fit to this distribution and the solid line shows the cut 
applied to identify short period variables.
\label{fig:f1}}
\end{figure}

For each object with at least 20 pairs of observations passing the tolerant
cuts, we calculate the modified Welch-Stetson \citep{wel93} variability index
\ival \space described in \citet{ake00a}. The distribution of variability indices 
seen in a representative sample of the data is shown in Figure 
\ref{fig:f1}. This distribution is roughly Gaussian with
a mean value of 0.227 and a width $\sigma$ = 0.11. Every object with \ival \space greater than one, about 7 $\sigma$ from the mean, is accepted as a variable.
From a list of 1.43x10$^7$ input light curves, 63665 are selected as variable by these criteria, a 
variability fraction of 0.45\%. The total number of detectable variables, 
as can be seen from the \ival \space distribution plot, is substantially larger.

Most of the variable objects we identify are long period variables, with
periods of 10 days or more. To identify short period variables within this 
set, we calculate a simple light curve roughness parameter similar
in spirit to the WS variability parameter. For this 
calculation we consider triplets of consecutive pairs of observations spaced by no more than five days. For
each of the outer pairs in a triplet we calculate the mean magnitude. Using these, we 
predict, by linear interpolation, the mean magnitude for the middle 
pair. We then compare the residual between actual and predicted magnitude
for each observation in the central pair to the error in its magnitude. 
\begin{eqnarray}
\delta_1 = \frac{m_1 - m_{predicted}}{\sqrt{\sigma_1^2 + 0.04^2}} \\
\delta_2 = \frac{m_2 - m_{predicted}}{\sqrt{\sigma_2^2 + 0.04^2}}
\end{eqnarray}
An additional uncertainty of 0.04 magnitudes is added in quadrature to the
measurement errors to reduce the sensitivity of this parameter to small
non-Gaussian errors. We then sum the absolute values of the products of 
these scaled residuals, divided by the number of predicted points, to 
construct our roughness parameter:
\begin{equation}
R = \frac{1}{\sqrt{N_{triplets}\times (N_{triplets}-1)}}
    \Sigma_{triplets} \sqrt{|\delta_1 \times \delta_2|}
\end{equation}
The distribution of this roughness parameter for all of the 63665 variables
is shown in Figure \ref{fig:f1}. For variables with
periods longer than a few days, this roughness parameter is distributed in a
approximately Gaussian manner, with a mean of 0.4 and a width $\sigma$ = 0.22.
To construct a list of candidate short period variables we select those with $R > 1.0$. 
The total number of such candidates is 17,508.

\begin{figure}
\plotone{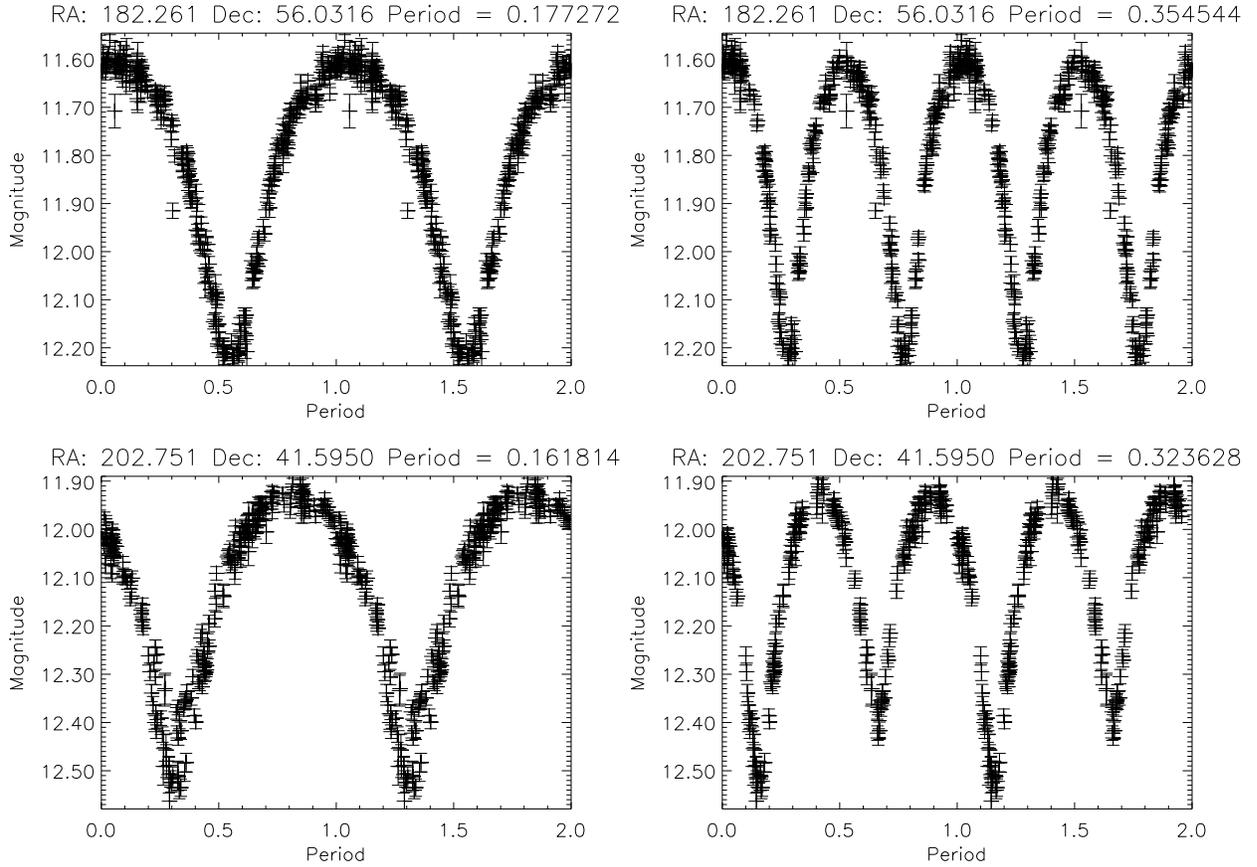}
\figcaption[figure2.eps]{Phased light curves for two
contact binary candidates. In cases where the light curve is fully symmetric,
as in the top two panels, it appears smooth and single-valued when phased 
with half the real period (left) or with the full period (right). In cases
where the minima are not identical, as in the bottom panels, the light curve
looks multi-valued and `rough' when phased at half the real period (left), 
but smooth when phased at the real period (right). In each case, data is repeated
to show two full cycles of the light curve.
\label{fig:f2}}
\end{figure}

Each of these candidate short period variables is then passed to the cubic
spline phasing code described in detail in \citet{ake94}. This code provides 
best-fit periods, period error estimates, and spline fit approximations
for light curve shapes. For each variable, we test the quality of this phasing
by measuring an analogous roughness parameter for the phased
light curve. For most eclipsing systems the period identified is actually
half the real period. If the light curve is symmetric (as is the case for 
full contact binaries), the phased light curve is very smooth with this period.
If the minima are not symmetric, the light curve will appear `rough' with
this period, but smooth when tested at twice the period. As a result we measure
light curve smoothness for both the identified period and for twice the
period, and accept as well phased light curves which have acceptably small
roughness with either period. Examples of this are shown in Figure 
\ref{fig:f2}.
Comparison of this folded roughness to the original
roughness allows us to define a sample which is well-phased. Of the 17,508
short period candidates, 16548 have a phased roughness parameter 
less than 1.5 when phased at either the derived period or twice the derived period. 
These objects are deemed to be confidently phased. 

\subsection{Combination of NSVS and 2MASS data}

ROTSE-I data is unfiltered, so although we have excellent light curve 
information, we have no color information. To ameliorate this, we combine
the ROTSE-I light curve information with J, H, and K band near-infrared data
drawn from the Two Micron All-Sky Survey (2MASS). This combination is 
especially apt because 2MASS data is significantly deeper than ROTSE-I
data. As a result 2MASS measurements for all ROTSE-I objects are 
rather precise. 

\begin{figure}
\epsscale{1}
\plotone{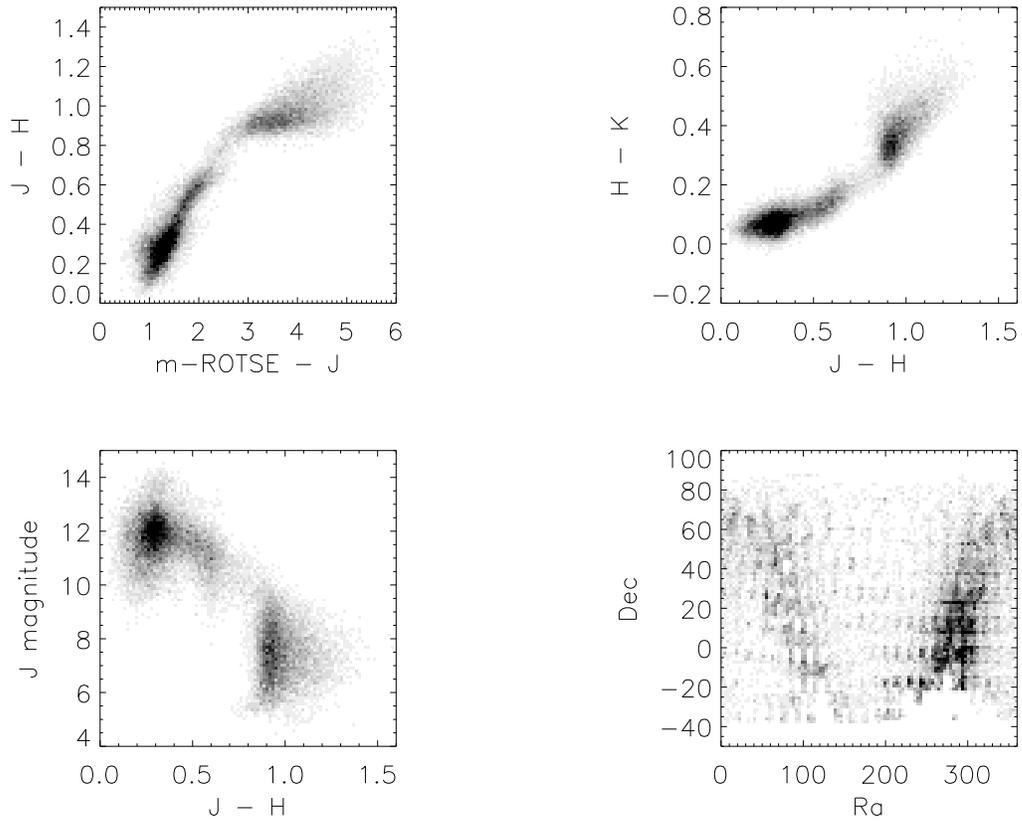}
\figcaption[figure3.eps]{Color-color plots 
in $m_{ROTSE} - J$, J - H, and H - K for all NSVS variables. The final panel 
shows an Aitoff
projection of all objects.
\label{fig:f3}}
\end{figure}

\begin{figure}
\epsscale{1}
\plotone{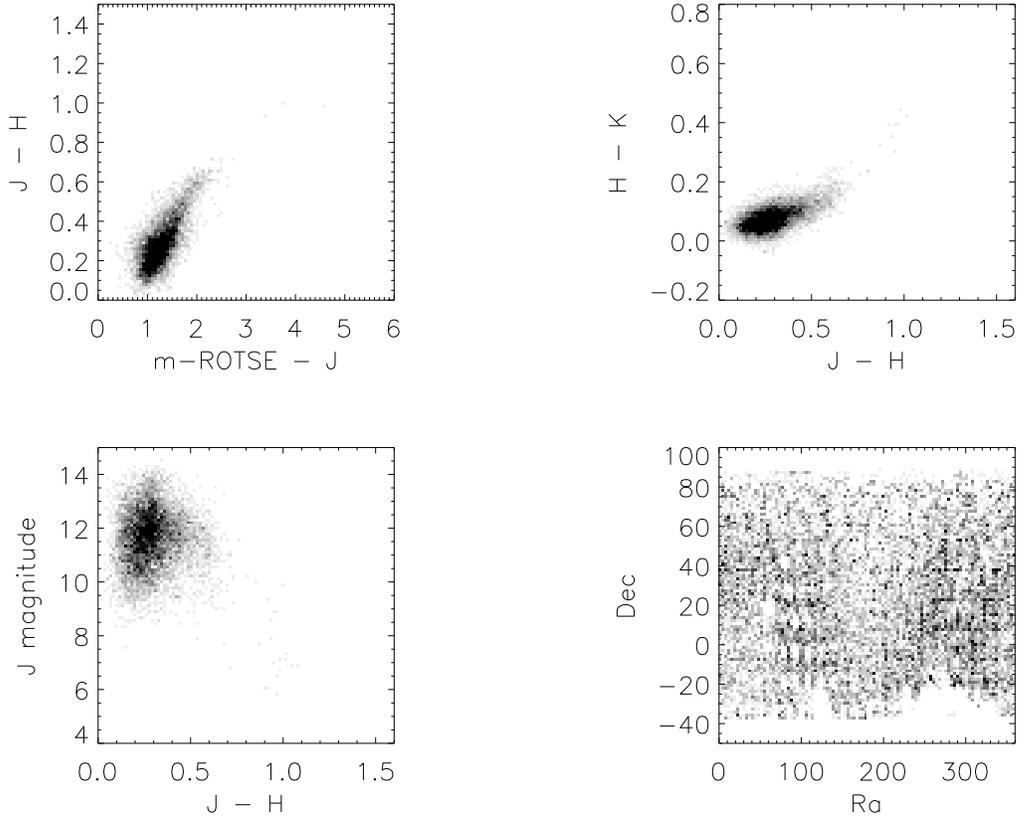}
\figcaption[figure4.eps]{Color-color plots 
in $m_{ROTSE} - J$, 
J - H, and H - K for those NSVS variables identified as potential short
period variables by their light curve roughness. The final panel shows an Aitoff
projection of all objects. It is clear from comparison to Figure \ref{fig:f3}
that the short period variables are drawn primarily from the blue population of stars.
\label{fig:f4}}
\end{figure}

2MASS observations are taken simultaneously in all three bands, and are 
reported for a single epoch.
Combining these data with ROTSE information provides three colors, 
m$_{ROTSE}$-J, J-H, and H-K, where m$_{ROTSE}$ is the mean apparent magnitude in 
the unfiltered ROTSE-I band. Color-color plots for all variables and for short period
variables are shown in Figures \ref{fig:f3} and 
\ref{fig:f4}.
The 2MASS observations correspond to random phases of the ROTSE light curves, 
generating uncertainty in each value of m$_{ROTSE}$-J and causing the horizontal 
spread in the m$_{ROTSE}$-J v.\ J-H  plots. The J-H and H-K color measurements do not suffer this 
dispersion, and the measured J, H, and K magnitudes can be simply compared 
to determine single epoch object colors.

To identify the proper 2MASS counterparts we pass the 
positions of all NSVS variables to the 2MASS database query tool at
IPAC\footnote{http://www.ipac.caltech.edu/2mass/}. Since ROTSE-I pixels
are relatively large, there can be minor ambiguity in the identification
of the correct corresponding 2MASS source. This problem is limited by the
fact that ROTSE-I variables are all rather bright and hence their sky 
density is not especially high. When there is more than one match, the 
chosen 2MASS matching object is the nearest object with 
m$_{ROTSE} - J > 0.$

Color-color plots for all the variables and those identified as short
period by the methods described above are shown in Figures 
\ref{fig:f3} and \ref{fig:f4}. It is clear by 
comparison that the short period candidates are a special, predominantly
blue, subset. Visual examination of the few red (m$_{ROTSE} - J > 3.0$)
short period candidates shows them to all be long period variables
with relatively large light curve roughness. As a result, we further refine
our short period variable candidate list by requiring 
m$_{ROTSE} - J \le 3.0$ and $H - K \le 0.35$. These cuts leave a total of
16046 phased short period variable candidates.

\subsection{Selection of Contact Binary Stars}

\begin{figure}
\epsscale{1}
\plotone{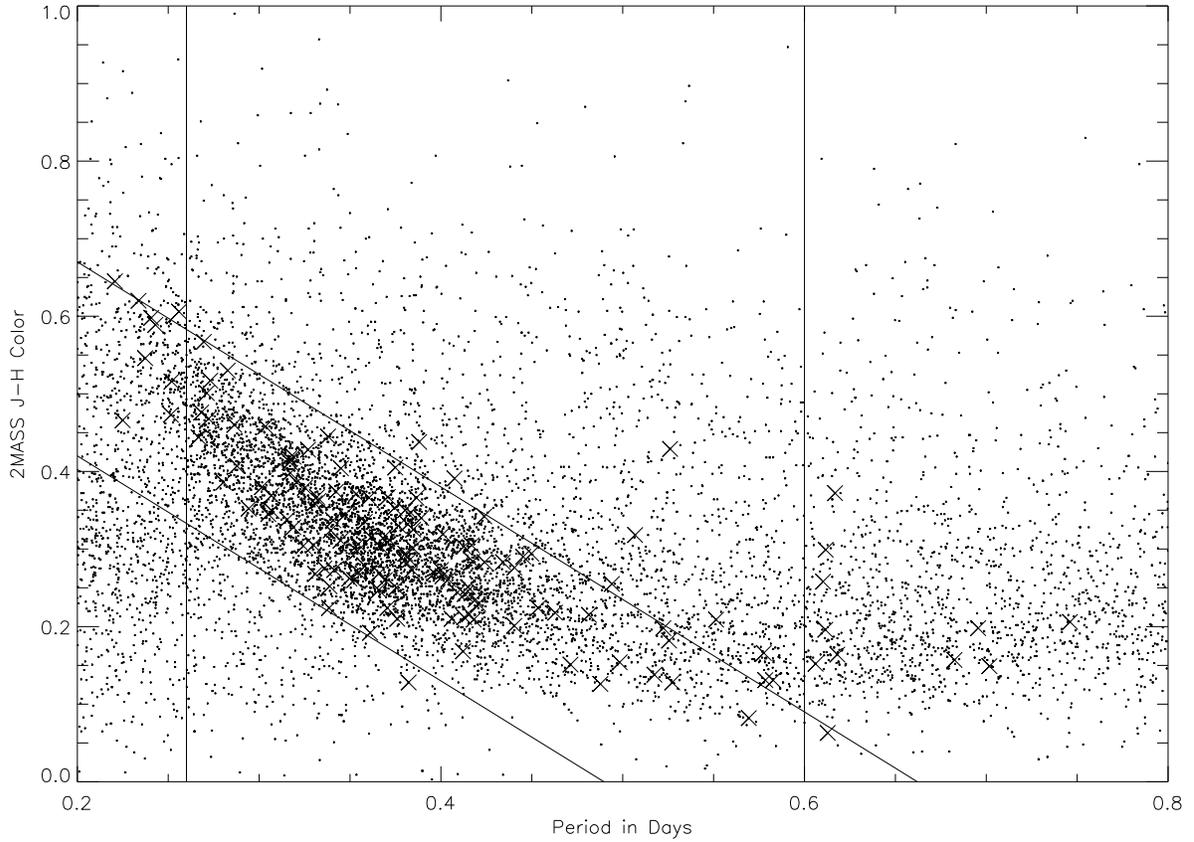}
\figcaption[figure5.eps]{Period-color
relation for the full short period variable sample (dots). A clear sequence of
objects with the period-color dependence expected for contact binary stars stands out.
We select as contact binary candidates all stars falling within the selection
region shown. The known contact binaries from the catalog of 
\citet{pri03} are also included (Xs).
\label{fig:f5}}
\end{figure}

Contact binary stars are known to exhibit a period-color relation. As shown in 
Figure \ref{fig:f5}, there is a dense patch of short period variables 
which displays a period-color dependence. Comparison to the location in period-color space of
known contact binaries suggests that this excess is largely due to these stars. 
To generate a list of potential contact binaries, cuts were made defining the region:
\begin{eqnarray}
0.26 < \Gamma < 0.6 \\
0.71 - 1.45\Gamma < J-H < 0.96 - 1.45\Gamma
\end{eqnarray}
selecting a total of 5179 objects. Here, the period ($\Gamma$) 
\rm is twice the value returned by the phasing procedure, and so represents the true period for 
contact binaries.

Interspersed throughout the region of period-color dependence is a background 
of variables which are not contact binaries. 
Upon scanning the light curves, many of the 5179 candidates were found to be 
other types of variables. In addition, many of the candidates with lower \ival\ 
parameters had small-amplitude light curves and large photometry errors, making 
classification difficult.

To ensure an essentially pure sample of contact binaries, further data quality cuts were made to select 
only the very well measured light curves. A cut at \ival $> 2$ eliminated 62\% of 
candidates, many of which had relatively indistinct light curves. An additional cut was placed on 
the roughness of the light curve when phased at the true period, requiring this to fall in 
a range from 0.25 to 0.75. This removed an additional 14\% of the candidates, mostly
variables which were clearly not contact binaries and so had light curves characterized by 
a different phased roughness value.  The remaining 1238 candidates were individually scanned and 66 
objects that did not appear to be contact binaries were removed. 9 of these 
appeared to be RR Lyrae stars and 3 appeared to be ordinary eclipsing binaries. The other removed 
light curves were not sufficiently well measured to be easily identified and many appeared to have 
asymmetric brightening and dimming phases. After these cuts, a sample of 1172 contact binaries remains.

Finally, due to the overlap in the fields, some of the objects appear two or more times in the 
catalog. In these cases we kept in the catalog data from the field that provided the most good 
observations. After discarding the duplicate objects, we are left with a catalog of 1022 contact binaries. 
Based on visual inspection and comparison to existing catalogs we expect this sample to be rather 
pure, with perhaps 5\% contamination of pulsating variables such as RRc stars. A sample of this catalog
is given in Table \ref{table:sample1}. The full catalog is available online.

\subsection{Tests of Selection Efficiency}

In this section we describe efforts to determine the selection efficiency 
for contact binaries. We test the catalog of \citet{pri03} for inclusion in our catalog. 
Due to the data quality cuts, it is expected that our catalog will be most complete for 
nearby objects, but it is not possible to use a volume-limited sample, as their catalog does not 
contain distance estimates. Instead, we use a magnitude-limited sample, restricting our analysis 
of completeness to objects brighter than 12.5 magnitudes. This corresponds roughly to the objects 
whose distance estimates placed them within 300 pcs. 
In the Pribulla catalog, there were 274 known contact binaries with 
maximum apparent magnitude values less than 12.5. A total of 148 of these were located 
within the NSVS survey region. Among these, 50 objects were included
in the final catalog, resulting in a total completeness level of about 34\%.
This low efficiency reflects the stringent cuts made to ensure a pure sample. 

The remaining objects were not included in the catalog for a variety of reasons. The 148 Pribulla 
stars that could reasonably have been observed were compared to the list of 16548 well-phased short period variables. 
There were 88 matches, about 25\% of which fall outside of the period-color cuts we have defined. Another 20\% do 
not pass the data quality cuts. The remaining 60 objects do not have enough good observations, either because they 
fall in regions of high stellar density (near the galactic plane) or because they are in regions less well observed
in the NSVS.

\section{ESTIMATING DISTANCES TO CONTACT BINARIES \label{sec:distances}}

In this section we describe the calibration of a near-infrared period, color, luminosity relation 
for contact binaries. We also discuss applying this relation to estimate distances to all of our
contact binaries.

\subsection{Determining $V_{max}$}

\begin{figure}
\epsscale{1}
\plotone{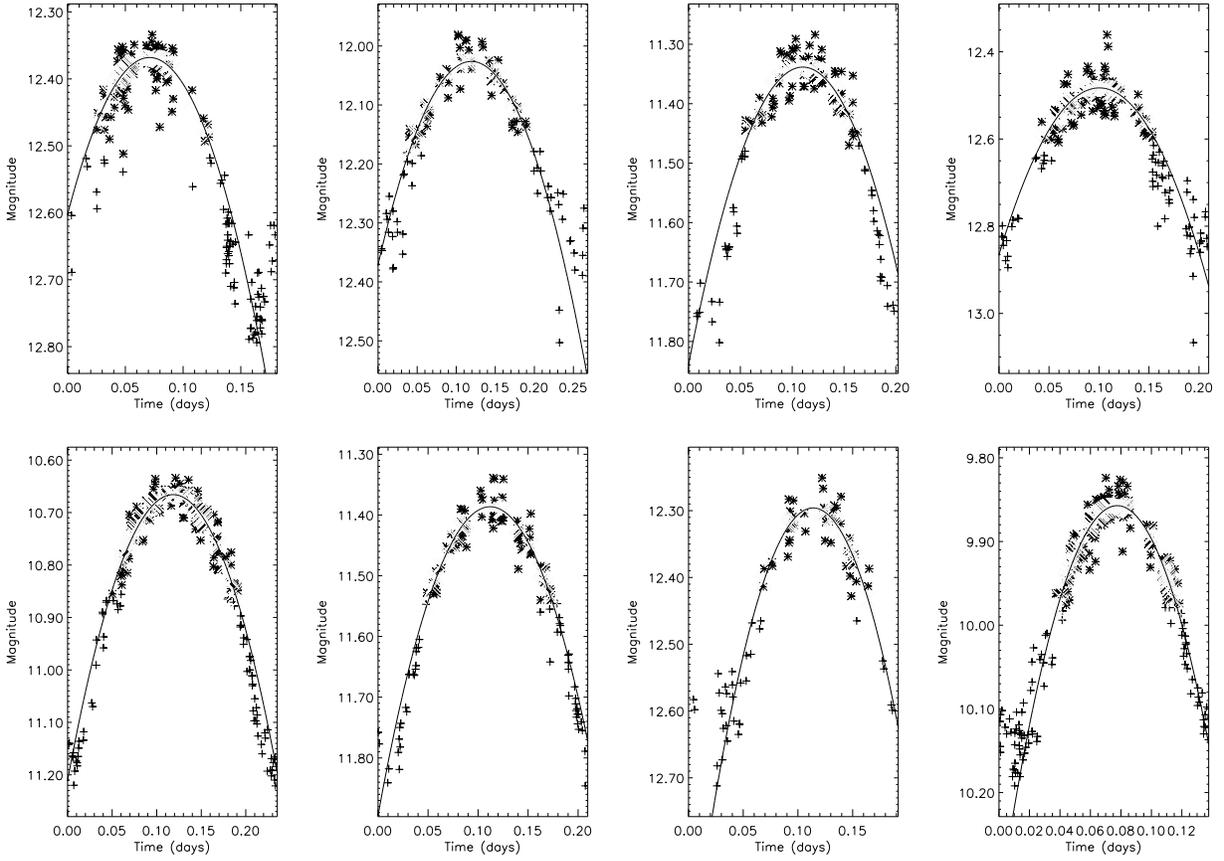}
\figcaption[figure6.eps]{Examples of the automatic
calculation of $V_{max}$ for a random set of eight contact binaries.
\label{fig:f6}}
\end{figure}

For contact binaries, the maximum magnitude $V_{max}$ is independent of
inclination, and hence is an appropriate measure of apparent magnitude for distance estimation.
The $V_{max}$ of each object was determined from its light curve. First, the light curves were phased, 
then the observations dimmer than the mean magnitude were removed and a parabola was fitted to the 
remaining points. The vertex of the parabola was taken to be the maximum magnitude. If the fit failed, 
$V_{max}$ was taken to be the average magnitude of the brightest observations. 
Figure \ref{fig:f6} shows the parabola fit used in obtaining $V_{max}$ for eight random 
objects in the catalog. 

\begin{figure}
\epsscale{1}
\plotone{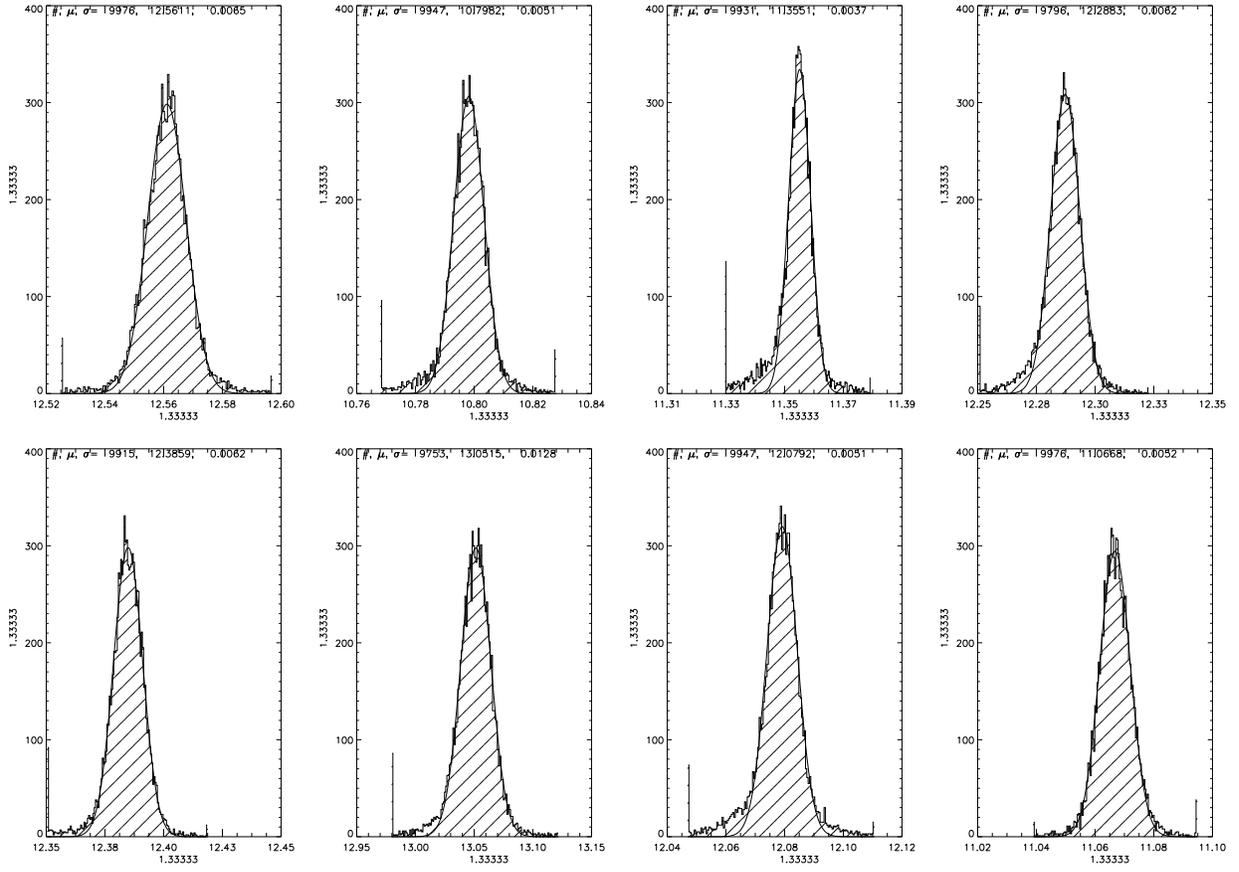}
\figcaption[figure7.eps]{Examples of the Gaussian fit to the distribution of bootstrap re-sampled $V_{max}$ values for a sample of eight contact binaries.
\label{fig:f7}}
\end{figure}

To obtain the error on $V_{max}$, this fit was bootstrap re-sampled
10000 times. Of the N datapoints for each star, a random sample of size 
N was selected, creating 
numerous slightly different fits from a single light curve. 
The resulting distribution of $V_{max}$ values 
was histogrammed using the optimal bin size and then fitted to a Gaussian. The mean of the Gaussian was 
taken to be $V_{max}$ and its standard deviation was taken to be the error on $V_{max}$. Examples of the 
Gaussian fit to the histogram are given in Figure \ref{fig:f7}.

For most objects, the mean of the Gaussian differed by less than .1 magnitude from the median of 
the bootstrap distribution. However, in a few the difference was much larger. These objects also 
had atypically large standard deviation for their Gaussian fits. In this case, $V_{max}$ values at 
4 $\sigma$ of the bootstrap distribution and beyond were thrown out, and the recalculated median 
and standard deviation were used as $V_{max}$ and its uncertainty, respectively. 

\subsection{Calibrating the Period-Color-Luminosity Relation}

To obtain distance estimates, a sample of 38 previously known contact binaries was used. Each of 
these objects has parallax data from the Hipparcos catalog, in addition to 2MASS color data and 
a period derived from its light curve. Reference distances were calculated from parallax, then 
three objects were removed due to poorly determined distance estimates. Absolute magnitude values
 were then determined from parallax. 

 \citet{ruc97} established a relation between the period, \bv\ 
color and absolute magnitude of a contact binary system. Contact binaries are close to the main 
sequence and so have a mass-radius dependence. Because the stars are in contact, the period of the 
system depends on the radii of the component stars, so the color of the system depends on its period. 

 We derived a similar relation using the {\sl J--H} color obtained by 
2MASS. A plane was fitted to the period, color and luminosity data of the 35 
calibration stars, which were weighted by distance uncertainty. The 
coefficients of this initial fit were sampled until a minimum value of 
$\chi^{2}$ was found and these new coefficients were selected to be the proper fit. The uncertainty on these 
coefficients was derived from their pattern of variation with increasing $\chi^{2}$. The observed relation can 
be written:
\begin{equation} 
M_{ROTSE}=(2.20\pm 0.66) + (0.88\pm 1.01)  \log(\Gamma) + (7.99\pm 2.33) (J-H).
\end{equation}

\begin{figure}
\epsscale{1}
\plotone{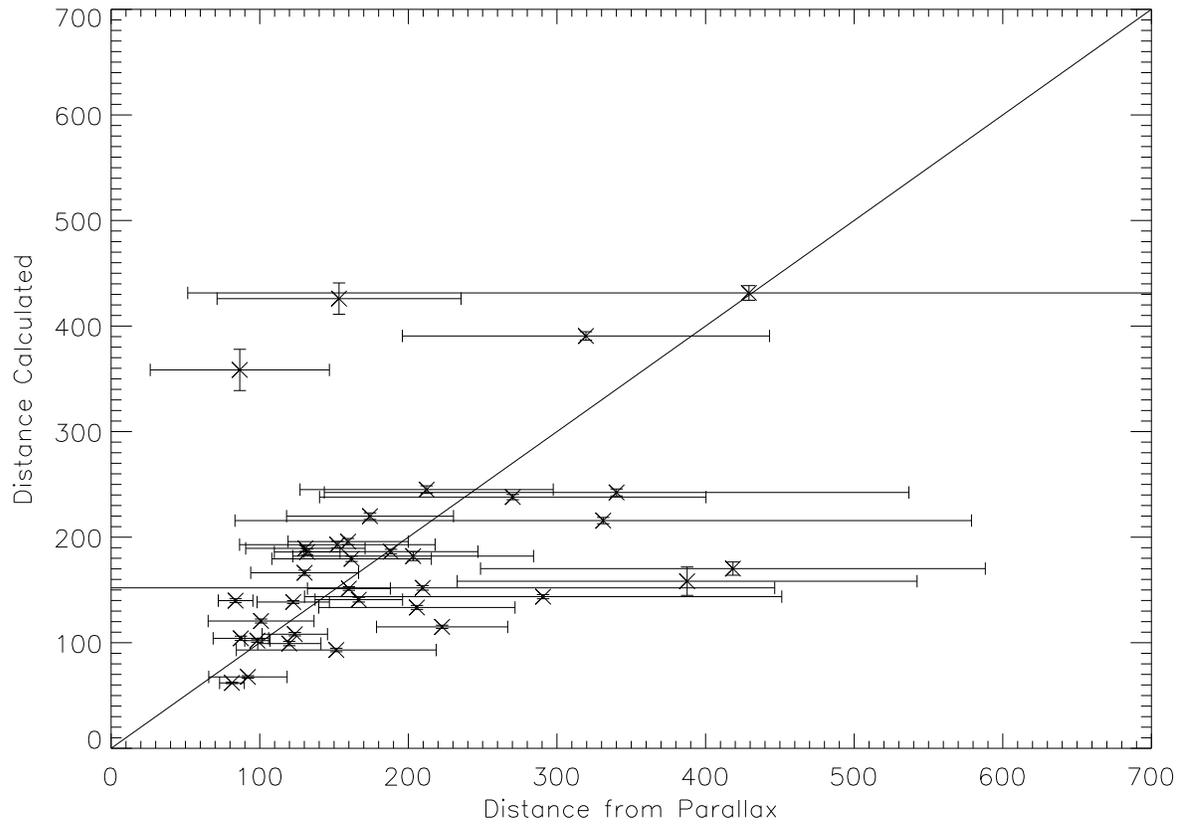}
\figcaption[figure8.eps]{Distances (pcs) from parallax for the reference set of 35 contact binaries, against the distances obtained from the period-color-luminosity relation. Also included is a reference line of slope = 1.
\label{fig:f8}}
\end{figure}

Used in combination with the standard magnitude-distance formulas, we obtain 
the relation
\begin{equation} 
\log(D)=0.2 V_{max} - 0.18 \log(\Gamma) - 1.60 (J-H) + 0.56
\end{equation}
where distance is in parsecs and \it{$\Gamma$} \rm is the true period in days. When used to calculate 
distances for our sample, all but about 23\% had values within 80 pcs of their distances from 
parallax. The median difference was 36 pcs and the maximum was 273 pcs. The comparison between 
the calculated distance and the distance from parallax is shown in Figure
\ref{fig:f8}. This relation was then used to obtain distance estimates for the full set of 
contact binaries.

\subsection{Comparison to Other Calibrations}

\begin{figure}
\epsscale{1}
\plotone{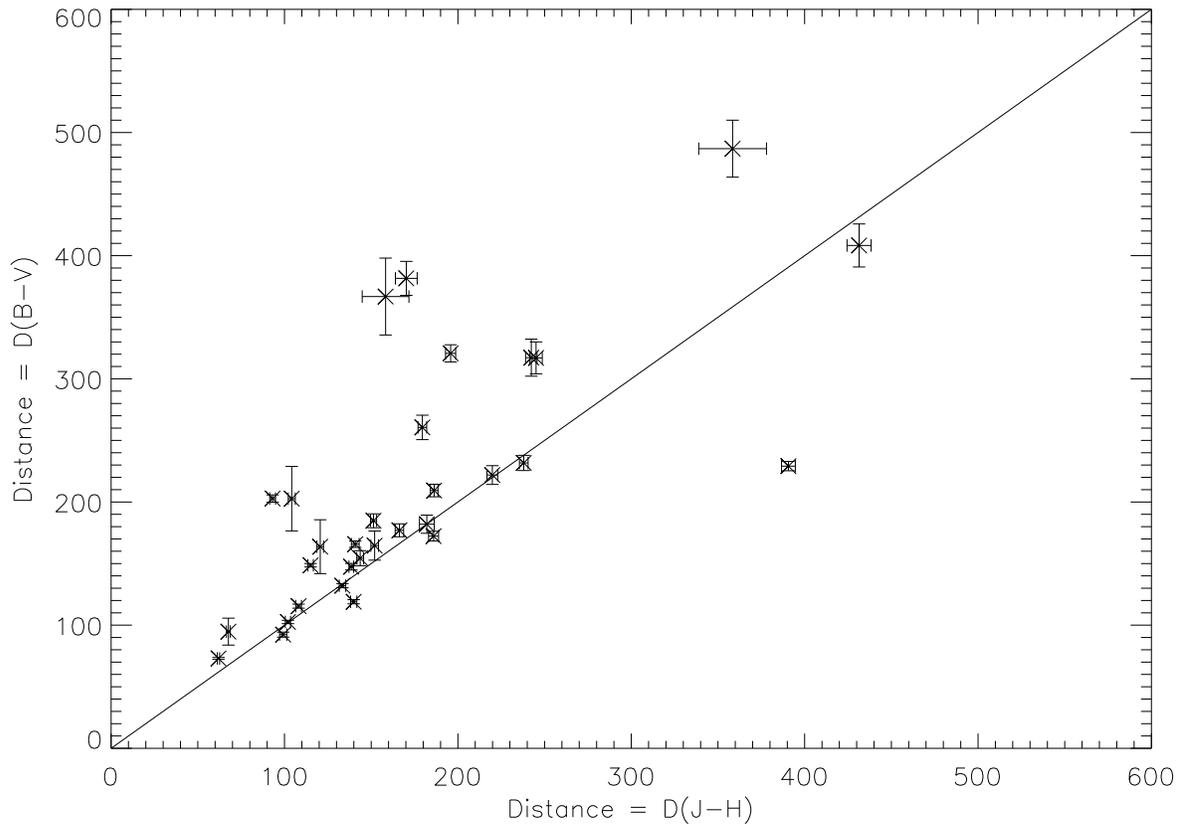}
\figcaption[figure9.eps]{Distances (pcs) obtained from our period-color-luminosity relation for the reference set contact binaries, against their distances from that of Rucinski, in \bv. Also included is a reference line of slope = 1.
\label{fig:f9}}
\end{figure}

Rucinski and Duerbeck's original period-color-luminosity relation was then applied to our calibration 
sample. Their absolute magnitude estimates tended to be lower and their distance estimates slightly 
higher than ours. See Figure \ref{fig:f9}. All but about 22\% of their distance 
estimates were within 80 pcs of the references distances. The median difference was 39 pcs and the maximum 
was 400 pcs; results which are comparable to those obtained here.
Rucinski and Duerbeck used a calibration sample of 40 systems, only slightly larger than ours. However, 
those systems have a larger range of period, $0.24 < \Gamma < 1.15$ days versus $0.24 < \Gamma < 1.06$,
and color, $0.26 < \bv < 1.14$ versus $0.28 < \bv < 0.87$, than do the 31 systems in our calibration
set for which \bv values are known. Therefore their calibration set may be more representative of the 
entire class of contact binaries.

\subsection{Tests for Third Parameters}
\begin{figure}
\epsscale{1.75}
\plottwo{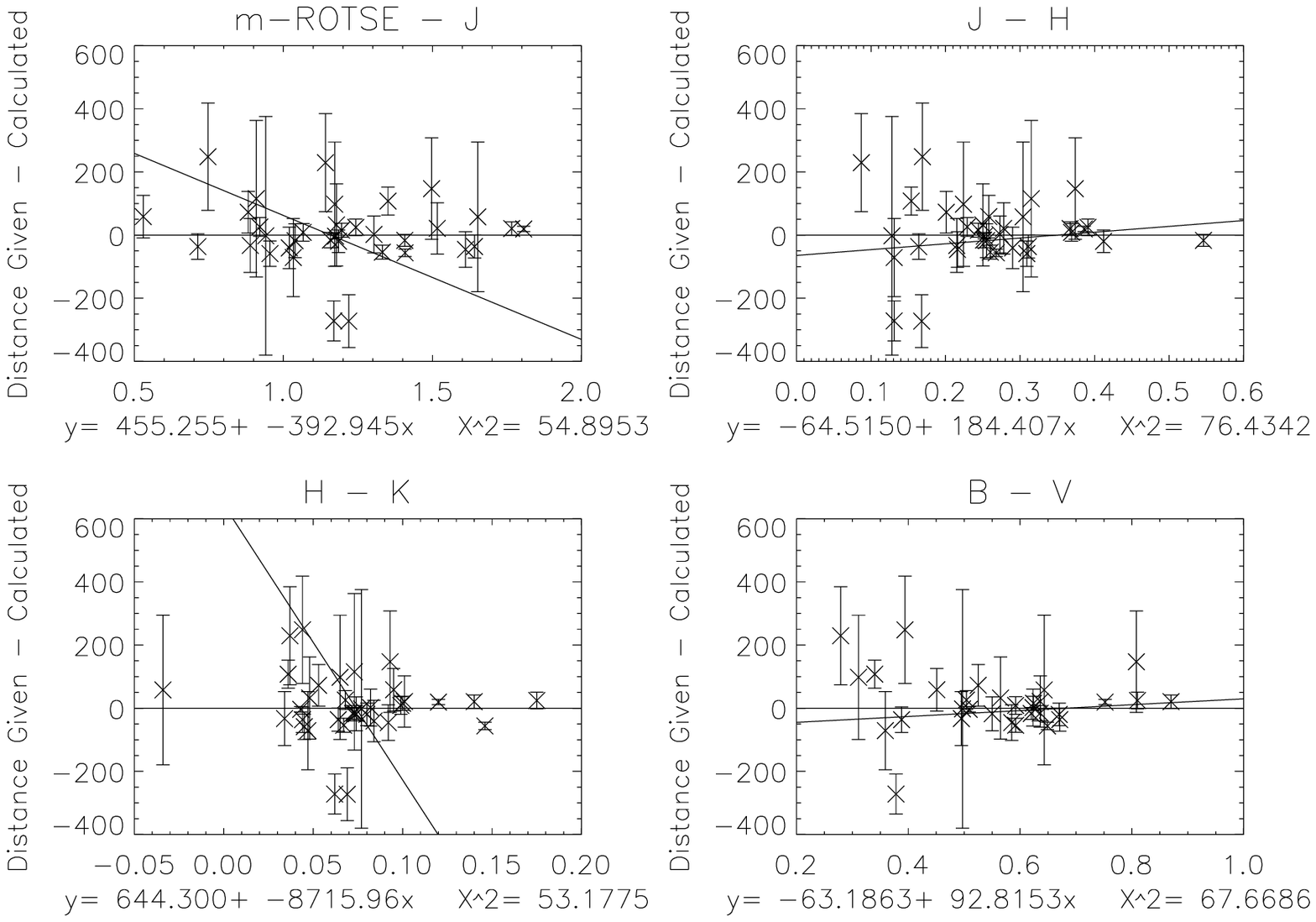}{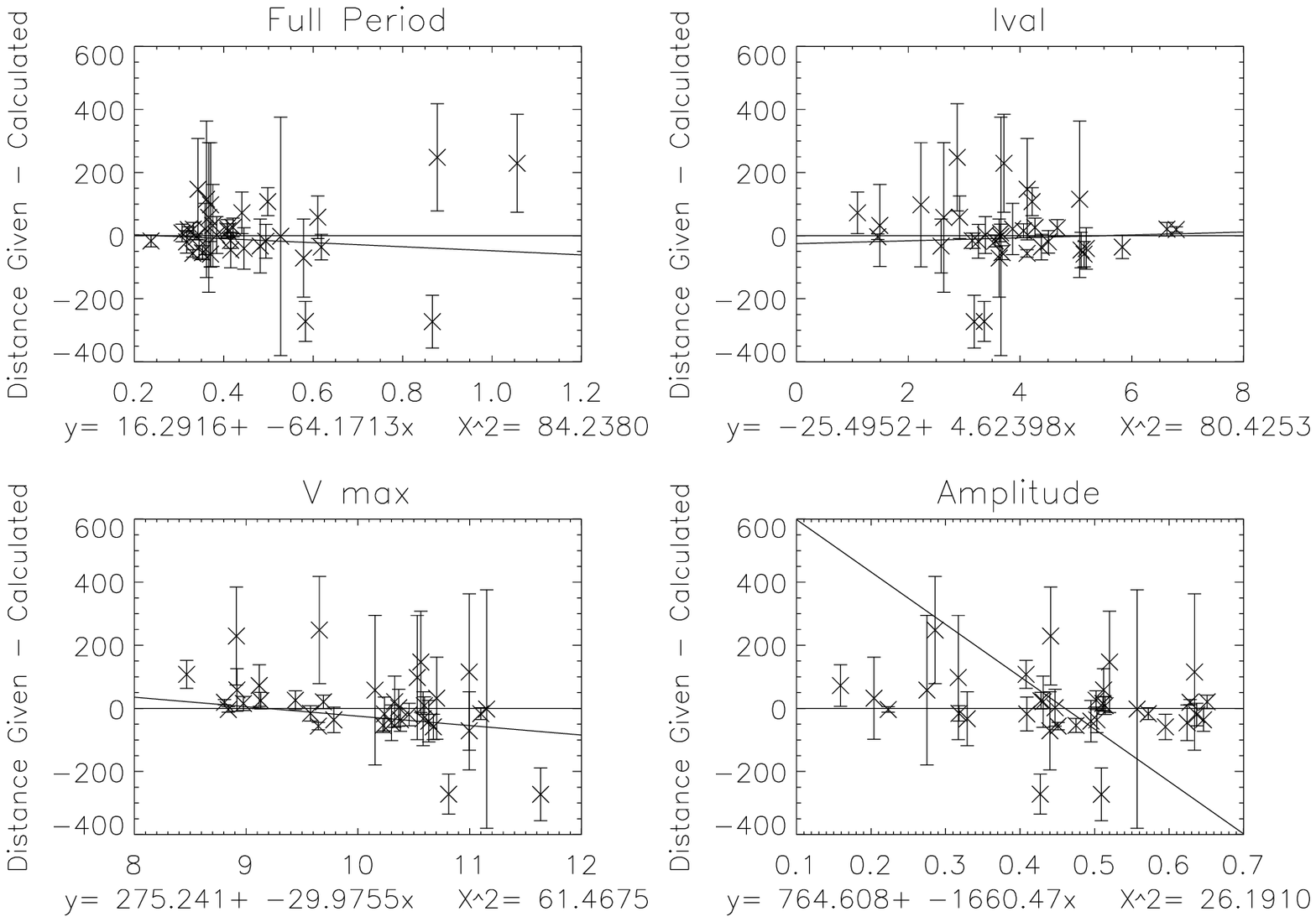}
\figcaption[figure10.eps]{Residuals in distance from our period-color-luminosity relation. The included linear fits do not seem indicative of any real trend.\label{fig:f10}}
\end{figure}

\begin{figure}
\epsscale{1.75}
\plottwo{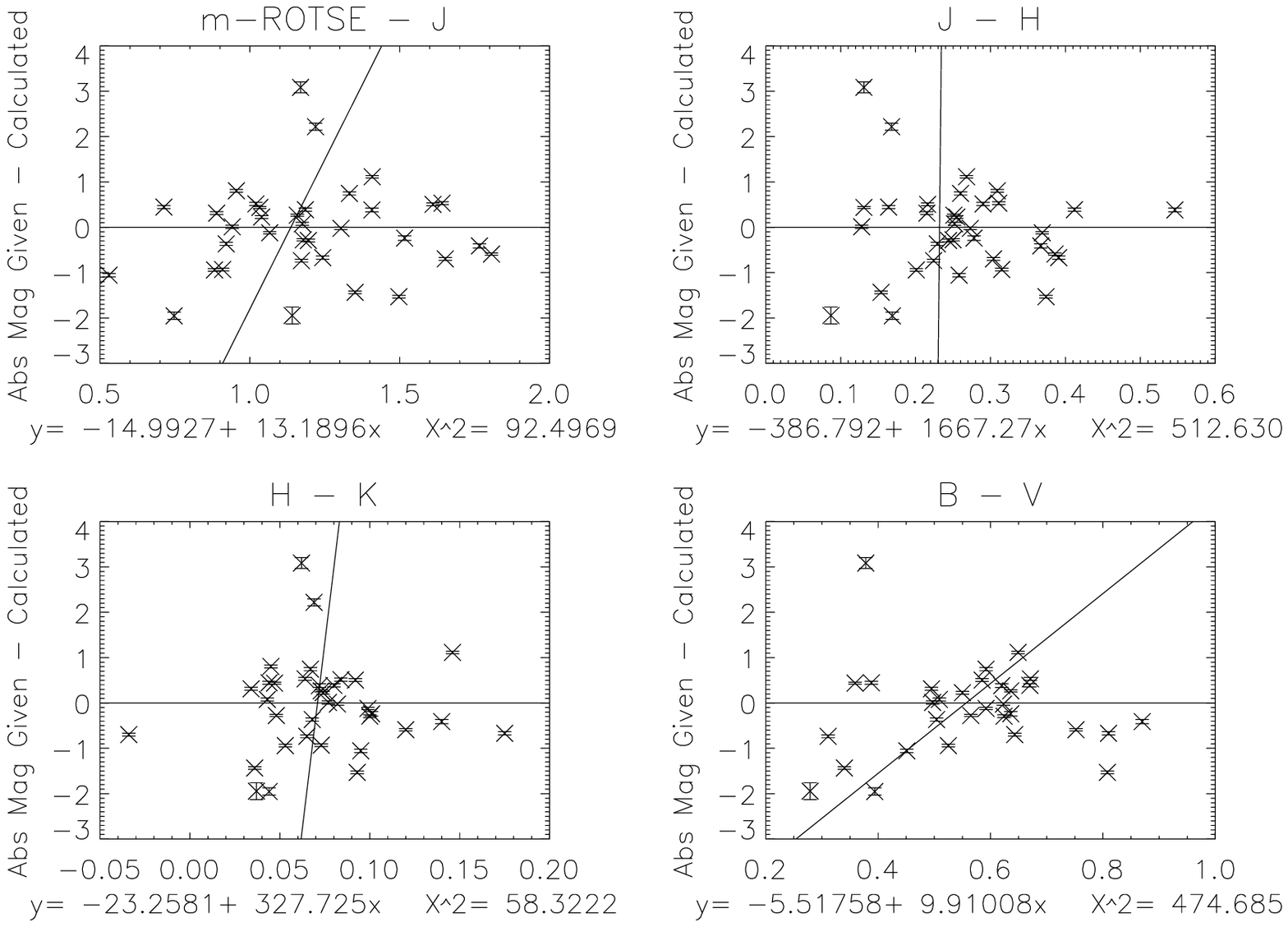}{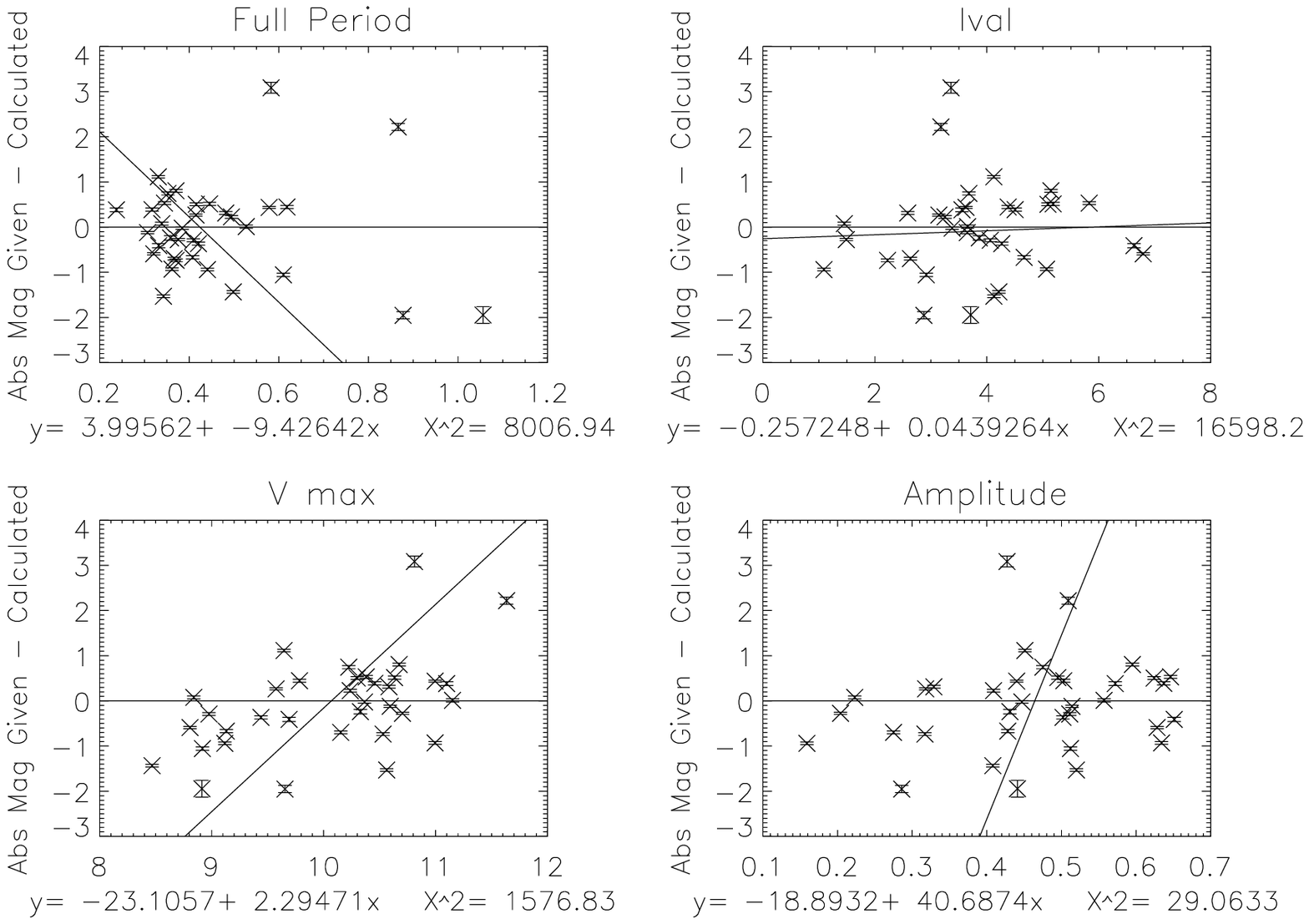}
\figcaption[figure12.eps]{Residuals in absolute magnitude from our period-color-luminosity relation. The included linear fits do not seem indicative of any real trend.\label{fig:f12}}
\end{figure}

Earlier absolute magnitude calibrations have included additional dependences, 
such as the metallicity and the orbital inclination of the system. To test for 
sources of dispersion in the period-color-luminosity relation, the distance residuals were plotted 
against all available colors: m$_{ROTSE}$-J, J-H, H-K, and \bv; as well as $\Gamma$, $V_{max}$, 
\ival, and the amplitude of the light 
curve. Any dependence on these last two parameters may be representative of a 
dependence on orbital inclination. Any relationship between the distance 
residuals and period or J-H would indicate that absolute magnitude depends more
strongly on those parameters than the calculated period-color-luminosity 
relation suggests. However, no significant dependencies were found. See plots in 
Figures \ref{fig:f10} and \ref{fig:f12}.

\section{ESTIMATING THE SPACE DENSITY OF CONTACT BINARIES \label{sec:space_density}}

The cumulative number of detections should increase with the distance to the sources cubed, if contact binaries 
are homogeneously distributed. To calculate 
the space density of contact binaries, we must first estimate the sky coverage of the contact binary catalog. We limit 
our analysis to the sky north of 0$^\circ$ declination, which was more thoroughly observed. This yields an area of 
17458 deg$^{2}$, or about 42\% of the sky. Therefore we take the total volume studied to be approximately $0.55\pi d^{3}$.
 
\begin{figure}
\epsscale{1}
\plotone{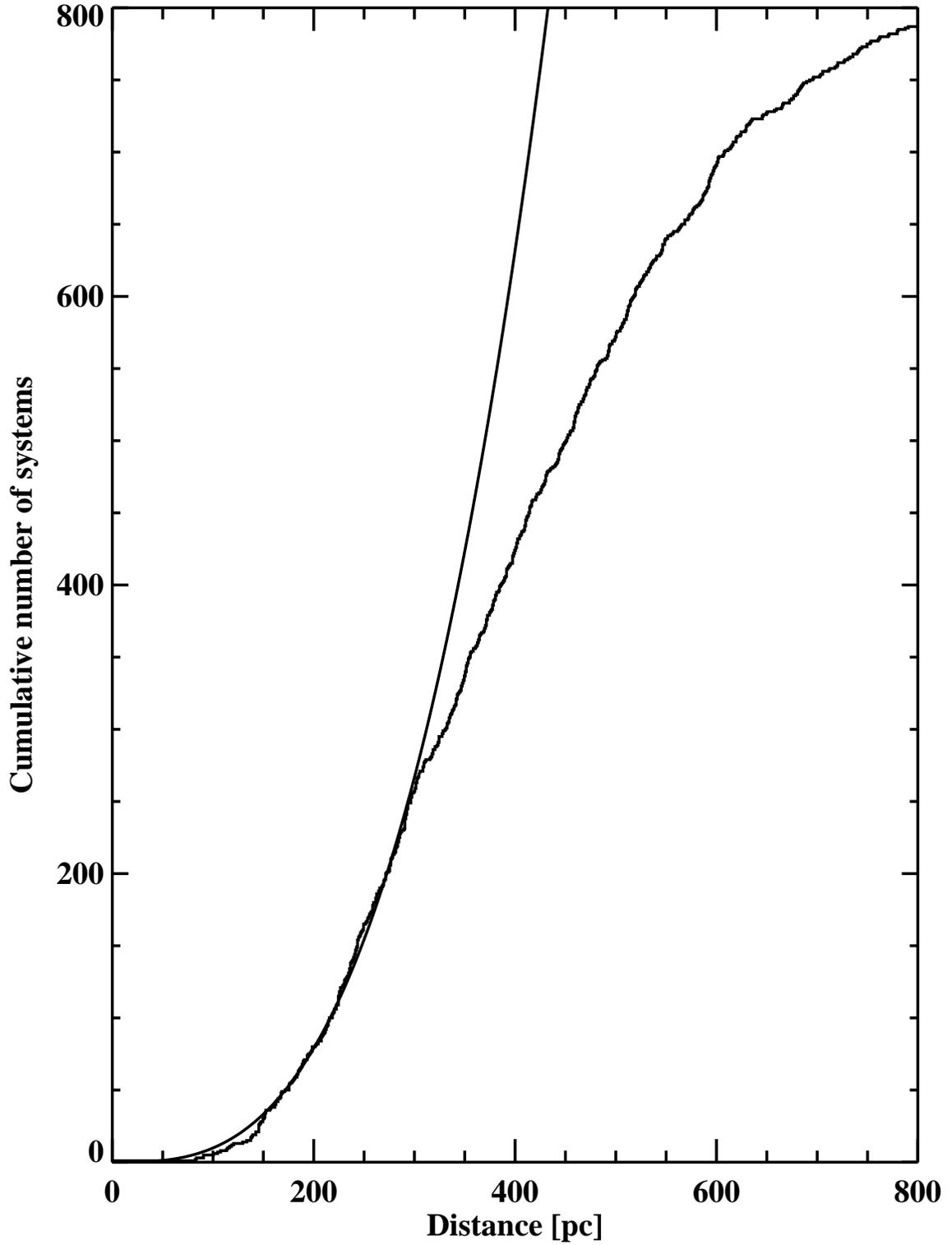}
\figcaption[figure14.eps]{Cumulative number of systems plotted as a function of distance. The dashed curve represents the curve $N = 9.9 \times 10^{-6}$d$^{3}$ systems per $pc^{3}$. The fit is very good out to about 300 pc.\label{fig:f14}}
\end{figure}

Using the distance estimates derived above and the method of \citet{ste01}, a curve was fit to the 
cumulative number of contact binaries detected as a function of distance, using only objects from 150 to 300 
parsecs. This is a distance range in which we expect uniform completeness. As can be seen in 
Figure \ref{fig:f14}, the curve $N = (9.9 \pm 3.7)\times 10^{-6}$d$^{3}$ gave a good fit over 
this distance range. This implies a measured space density of about $(5.7 \pm 2.1)\times 10^{-6} pcs^{-3}$. However, the catalog is only about 34\% complete for objects brighter than 12.5 magnitudes, which correspond approximately to those objects that are closer than 300 pcs, and thus used to determine the space density.  
Adjusting for this incompleteness, we obtain a space density of $(1.7 \pm 0.6) \times 10^{-5} pcs^{-3}$. This agrees well with the recent estimate by 
\citet{ruc02}, which was $(1.02 \pm 0.24) \times 10^{-5} pcs^{-3}$. 

\section{CONCLUSION \label{sec:conclusions}}

In this work we present a new catalog of 1022 bright contact binary stars.
All objects are selected from the extensive light curve database assembled in
the Northern Sky Variability Survey from data taken from ROTSE-I sky patrol
observations. Period, amplitude, light curve shape, and infrared colors are
all used to identify contact binary candidates. We also present a period-color-relation 
using {\sl J--H} and an estimate of the space density of contact binaries.
The detection efficiency for contact binaries given the stringent set of cuts applied
here was rather low, 34\%. This suggests that as many as a few thousand more contact 
binaries remain to be extracted from the NSVS data set. In addition, tens of thousands 
of additional variable stars of all kinds remain to be extracted from this powerful resource. 

\clearpage


\acknowledgments

This publication 
makes use of the data from the Northern Sky Variability Survey created jointly by the 
Los Alamos National Laboratory and University of Michigan. The NSVS was funded by the 
Department of Energy, the National Aeronautics and Space Administration and the National Science Foundation. 

This publication also makes use of data products from the Two Micron All Sky 
Survey, which is a joint project of the University of Massachusetts and the 
Infrared Processing and Analysis Center/California Institute of Technology, 
funded by the National Aeronautics and Space Administration and the National 
Science Foundation. 

ROTSE is supported at the University of Michigan by NSF grants AST 99-70818 and AST 04-07061, 
NASA grant NAG 5-5101, the Research Corporation, the University of Michigan and the Planetary Society. 
Gettel and Geske also acknowledge support from the University of Michigan REU site grant PHY 04-53355.
Gettel is also supported by a Michigan Space Grant Consortium undergraduate research grant.





\clearpage



\clearpage

\clearpage
\thispagestyle{empty}
\setlength{\voffset}{2.5cm}
\begin{deluxetable}{rcrcccccccccccccccccccccccll}
\rotate
\tablewidth{0pc}
\tablecolumns{28}
\tabletypesize{\tiny}
\tablecaption{Properties of sample objects in the contact binary catalog \label{table:sample1} \tablenotemark{a}}
\setlength{\tabcolsep}{0.03in}

\tablehead{\colhead{ID}  &  \colhead{Ra}  &  \colhead{Dec}  &  \colhead{m$_{R}$}  &  \colhead{m$_{R}$ err}  &  \colhead{J}  &  \colhead{H}  &  \colhead{K}  &  \colhead{J err}  &  \colhead{H err}  &  \colhead{K err}  &  \colhead{m$_{R}$-J}  &  \colhead{J-H}  &  \colhead{H-K}  &  \colhead{$\Gamma$} &  \colhead{\vm}  &  \colhead{\vm err}  &  \colhead{Max}  &  \colhead{Min}  &  \colhead{Amp}  &  \colhead{M}  &  \colhead{M err}  &  \colhead{D}  &  \colhead{D err}  &  \colhead{Field}  &  \colhead{Obs}  &  \colhead{GCVS Name}  &  \colhead{Ref}}
\startdata
8711652 & 321.925 & 35.0875 & 12.659 & 0.27 & 11.179 & 10.84 & 10.747 & 0.017 & 0.016 & 0.006 & 1.48 & 0.339 & 0.093 & 0.334968 & 12.5618 & 0.0055 & 12.523 & 13.459 & 0.936 & 4.48844 & 0.025072 & 412 & 5 &  066a  & 319 &               &            \\
7604549 & 176.101 & 23.3565 & 12.092 & 0.266 & 10.295 & 9.837 & 9.726 & 0.018 & 0.018 & 0.014 & 1.797 & 0.458 & 0.111 & 0.303427 & 11.9309 & 0.00524 & 11.907 & 12.716 & 0.809 & 5.40103 & 0.026801 & 202 & 3 &  057b  & 279 &     CE Leo *  &      GC,PR \\
13267072 & 201.887 & 3.04122 & 11.212 & 0.261 & 9.693 & 9.396 & 9.28 & 0.018 & 0.021 & 0.02 & 1.519 & 0.297 & 0.116 & 0.353995 & 11.0426 & 0.004465 & 11.029 & 11.754 & 0.724 & 4.17416 & 0.029287 & 236 & 3 &  105d  & 213 &     AW Vir *  &      GC,PR \\
7726255 & 211.242 & 30.0006 & 11.619 & 0.285 & 10.306 & 10.002 & 9.934 & 0.02 & 0.023 & 0.015 & 1.313 & 0.304 & 0.067 & 0.324288 & 11.5106 & 0.003869 & 11.498 & 12.224 & 0.726 & 4.19645 & 0.031747 & 290 & 4 &  059b  & 212 &     TU Boo *  &   GC,PR,RT \\
14850169 & 53.9278 & -20.352 & 11.847 & 0.271 & 10.233 & 9.898 & 9.845 & 0.019 & 0.02 & 0.019 & 1.614 & 0.335 & 0.052 & 0.327091 & 11.6192 & 0.007031 & 11.631 & 12.26 & 0.629 & 4.44737 & 0.029001 & 272 & 4 &  120b  & 188 &               &            \\
12142698 & 54.9961 & 3.24179 & 11.43 & 0.285 & 10.359 & 9.886 & 9.735 & 0.021 & 0.022 & 0.017 & 1.071 & 0.473 & 0.151 & 0.282723 & 11.3186 & 0.004425 & 11.307 & 12.067 & 0.76 & 5.49374 & 0.031427 & 146 & 2 &  096a  & 262 &               &            \\
12395130 & 87.4975 & -7.4318 & 12.47 & 0.3 & 11.114 & 10.662 & 10.533 & 0.021 & 0.022 & 0.019 & 1.356 & 0.452 & 0.129 & 0.322597 & 12.305 & 0.006254 & 12.274 & 13.167 & 0.893 & 5.37661 & 0.031672 & 243 & 4 &  098b  & 243 &               &            \\
12310373 & 77.2155 & 2.82084 & 11.418 & 0.241 & 9.77 & 9.339 & 9.272 & 0.023 & 0.02 & 0.019 & 1.648 & 0.431 & 0.066 & 0.295713 & 11.2387 & 0.00477 & 11.236 & 11.88 & 0.644 & 5.17549 & 0.031565 & 163 & 2 &  097d  & 246 &               &            \\
7655121 & 188.771 & 23.3371 & 10.746 & 0.276 & 9.326 & 9.102 & 9.002 & 0.018 & 0.019 & 0.015 & 1.42 & 0.224 & 0.1 & 0.338505 & 10.5973 & 0.004076 & 10.594 & 11.257 & 0.663 & 3.5739 & 0.027753 & 254 & 3 &  058b  & 292 &     RZ Com *  &   GC,PR,RT \\
706127 & 109.27 & 77.1739 & 11.676 & 0.247 & 10.39 & 9.974 & 9.884 & 0.021 & 0.017 & 0.013 & 1.286 & 0.416 & 0.09 & 0.298445 & 11.5549 & 0.006201 & 11.536 & 12.23 & 0.693 & 5.05921 & 0.028256 & 199 & 3 &  005d  & 691 &               &            \\
9179857 & 17.3828 & 22.6553 & 11.396 & 0.254 & 10.303 & 10.13 & 10.077 & 0.019 & 0.02 & 0.018 & 1.093 & 0.173 & 0.053 & 0.494339 & 11.2792 & 0.004494 & 11.263 & 11.932 & 0.669 & 3.31178 & 0.031024 & 392 & 6 &  069d  & 318 &               &            \\
7463810 & 144.789 & 31.4129 & 12.15 & 0.248 & 10.624 & 10.314 & 10.235 & 0.02 & 0.021 & 0.016 & 1.526 & 0.309 & 0.079 & 0.358664 & 12.019 & 0.005425 & 12.014 & 12.645 & 0.631 & 4.28301 & 0.030596 & 353 & 5 &  055a  & 279 &               &            \\
1119084 & 258.558 & 76.704 & 11.255 & 0.245 & 9.541 & 9.133 & 9.033 & 0.018 & 0.014 & 0.013 & 1.714 & 0.408 & 0.1 & 0.325397 & 11.1248 & 0.004801 & 11.099 & 11.953 & 0.854 & 5.02846 & 0.024481 & 166 & 2 &  009d  & 689 &               &            \\
9216203 & 22.5684 & 13.557 & 11.18 & 0.251 & 9.659 & 9.289 & 9.215 & 0.018 & 0.031 & 0.016 & 1.521 & 0.37 & 0.073 & 0.324736 & 11.0303 & 0.003985 & 11.016 & 11.708 & 0.692 & 4.72416 & 0.036933 & 182 & 3 &  070b  & 295 &               &            \\
14887343 & 62.9492 & -11.79 & 11.565 & 0.25 & 10.144 & 9.839 & 9.747 & 0.025 & 0.02 & 0.019 & 1.421 & 0.305 & 0.092 & 0.416889 & 11.4368 & 0.005135 & 11.423 & 12.148 & 0.724 & 4.30078 & 0.033995 & 267 & 4 &  120d  & 214 &     BL Eri *  &      GC,PR \\
269914 & 26.4939 & 80.0821 & 11.295 & 0.258 & 10.297 & 9.969 & 9.923 & 0.019 & 0.027 & 0.019 & 0.998 & 0.328 & 0.045 & 0.318828 & 11.1523 & 0.002926 & 11.131 & 11.785 & 0.654 & 4.38164 & 0.034154 & 226 & 4 &  002d  & 725 &     GW Cep *  &      GC,PR \\
13334244 & 213.254 & 6.94132 & 12.733 & 0.27 & 11.411 & 11.083 & 11.021 & 0.021 & 0.025 & 0.023 & 1.322 & 0.328 & 0.062 & 0.333536 & 12.6732 & 0.008905 & 12.651 & 13.287 & 0.636 & 4.39894 & 0.033896 & 452 & 7 &  106d  & 200 &               &            \\
12106392 & 47.4697 & -6.8931 & 10.766 & 0.234 & 9.746 & 9.456 & 9.372 & 0.021 & 0.023 & 0.021 & 1.02 & 0.29 & 0.084 & 0.445283 & 10.6332 & 0.004967 & 10.639 & 11.134 & 0.495 & 4.20624 & 0.033508 & 193 & 3 &  095c  & 132 &     UX Eri *  &      GC,PR \\
14086961 & 293.851 & 5.83833 & 10.806 & 0.239 & 9.851 & 9.542 & 9.497 & 0.02 & 0.021 & 0.021 & 0.955 & 0.309 & 0.045 & 0.370319 & 10.6743 & 0.007594 & 10.647 & 11.242 & 0.594 & 4.2873 & 0.030697 & 189 & 3 &  112a  & 214 &  V0417 Aql *  &      GC,PR \\
7661848 & 192.84 & 27.2296 & 12.219 & 0.25 & 10.892 & 10.448 & 10.332 & 0.019 & 0.031 & 0.017 & 1.327 & 0.444 & 0.116 & 0.266683 & 12.0274 & 0.004223 & 12.021 & 12.595 & 0.574 & 5.2397 & 0.037138 & 228 & 4 &  058b  & 236 &       EK Com  &   GC,PR,RT \\
9233832 & 31.6594 & 14.2572 & 11.115 & 0.247 & 9.742 & 9.51 & 9.466 & 0.015 & 0.021 & 0.021 & 1.373 & 0.231 & 0.044 & 0.484974 & 10.9769 & 0.008783 & 10.992 & 11.506 & 0.514 & 3.7757 & 0.029274 & 276 & 4 &  070c  & 145 &               &            \\
8019900 & 267.723 & 29.8529 & 11.732 & 0.279 & 10.425 & 10.223 & 10.2 & 0.018 & 0.027 & 0.02 & 1.307 & 0.202 & 0.022 & 0.434212 & 11.5712 & 0.005143 & 11.55 & 12.272 & 0.722 & 3.49368 & 0.034593 & 413 & 7 &  062c  & 225 &               &         RT \\
9447688 & 69.1569 & 18.755 & 10.506 & 0.256 & 8.895 & 8.679 & 8.587 & 0.017 & 0.021 & 0.018 & 1.611 & 0.216 & 0.092 & 0.41568 & 10.2997 & 0.008888 & 10.3 & 10.924 & 0.624 & 3.58878 & 0.029322 & 220 & 3 &  073a  & 295 &     RZ Tau *  &      GC,PR \\
11768183 & 335.737 & 16.3244 & 11.124 & 0.258 & 10.215 & 9.9 & 9.827 & 0.019 & 0.02 & 0.015 & 0.908 & 0.315 & 0.073 & 0.361483 & 10.9953 & 0.005459 & 10.967 & 11.601 & 0.634 & 4.32596 & 0.029285 & 216 & 3 &  090d  & 302 &     BB Peg *  &      GC,PR \\
10653145 & 241.842 & 10.4943 & 12.108 & 0.231 & 10.881 & 10.473 & 10.247 & 0.021 & 0.029 & 0.019 & 1.227 & 0.408 & 0.226 & 0.350103 & 11.9919 & 0.008208 & 11.969 & 12.769 & 0.8 & 5.05653 & 0.037051 & 244 & 4 &  084c  & 227 &               &            \\
6734014 & 56.7999 & 25.1164 & 11.498 & 0.17 & 10.279 & 9.917 & 9.899 & 0.021 & 0.029 & 0.018 & 1.219 & 0.362 & 0.017 & 0.33266 & 11.4191 & 0.005062 & 11.411 & 12.111 & 0.7 & 4.66951 & 0.03694 & 224 & 4 &  050b  & 330 &     AH Tau *  &      GC,PR \\
890396 & 187.889 & 68.6355 & 12.178 & 0.255 & 10.451 & 9.97 & 9.884 & 0.016 & 0.015 & 0.015 & 1.727 & 0.481 & 0.086 & 0.279921 & 12.0475 & 0.00474 & 12.037 & 12.821 & 0.784 & 5.55382 & 0.023296 & 199 & 2 &  007c  & 573 &               &            \\
6715004 & 61.2029 & 33.9559 & 11.342 & 0.212 & 9.934 & 9.619 & 9.54 & 0.023 & 0.032 & 0.016 & 1.408 & 0.315 & 0.079 & 0.373044 & 11.2206 & 0.0056 & 11.215 & 11.805 & 0.59 & 4.33803 & 0.040692 & 238 & 5 &  050a  & 373 &               &            \\
5189081 & 234.205 & 47.622 & 12.633 & 0.257 & 11.606 & 11.198 & 11.117 & 0.013 & 0.014 & 0.018 & 1.027 & 0.408 & 0.081 & 0.36047 & 12.4974 & 0.003901 & 12.471 & 13.12 & 0.649 & 5.06772 & 0.021472 & 306 & 3 &  039d  & 371 &               &            \\
15956050 & 185.328 & -13.997 & 11.304 & 0.206 & 10.147 & 9.951 & 9.918 & 0.018 & 0.021 & 0.019 & 1.157 & 0.196 & 0.033 & 0.474492 & 11.1662 & 0.004629 & 11.176 & 11.619 & 0.443 & 3.47978 & 0.03074 & 345 & 5 &  128d  & 119 &               &            \\
\enddata
\tablenotetext{a}{Ra and Dec information are taken from the corresponding 2MASS observations, due to the better spacial resolution. m$_{R}$ is the apparent ROTSE magnitude and $\Gamma$ is given in days. Max is the average magnitude of the brightest observations and Min is the average magnitude of the dimmest. Amp is the difference between them. D is given in parsecs and Obs is the number of observations passing our loose set of cuts. Included references are to the GCVS, the Pribulla catalog and the ROTSE-I test fields.}
\end{deluxetable}

\end{document}